\documentclass[a4paper]{PoS}

\usepackage[subrefformat=parens]{subcaption}
\title{The calibration of the first Large-Sized Telescope of the Cherenkov Telescope Array}

\ShortTitle{The calibration of the LST1}

\author{
\scriptsize
\speaker{S.~Sakurai}$^a$
, D.~Depaoli$^b$
, R.~L\'{o}pez-Coto$^c$,
J.~Becerra~Gonz\'{a}lez$^d$, A.~Berti$^b$, O.~Blanch$^e$,
F.~Cassol$^f$, A.~Chiavassa$^b$, D.~Corti$^c$,
A.~De~Angelis$^c$, C.~Delgado$^g$, C.~D\'{i}az$^g$, F.~Di~Pierro$^b$, L.~Di~Venere$^h$, M.~Doro$^c$,
A.~Fern\'{a}ndez-Barral$^e$,
F.~Giordano$^h$, S.~Griffiths$^e$,
D.~Hadasch$^a$,
Y.~Inome$^a$,
L.~Jouvin$^e$,
D.~Kerszberg$^e$, H.~Kubo$^i$,
A.~L\'{o}pez-Oramas$^d$,
M.~Mallamaci$^c$, M.~Mariotti$^c$, G.~Mart\'{i}nez$^g$, S.~Masuda$^i$, D.~Mazin$^{aj}$, A.~Moralejo$^e$, E.~Moretti$^e$,
T.~Nagayoshi$^k$, D.~Ninci$^e$, L.~Nogu\'{e}s$^e$, S.~Nozaki$^i$,
A.~Okumura$^l$,
R.~Paoletti$^m$, P.~Penil$^n$, R.~Pillera$^h$, C.~Pio$^e$
R.~Rando$^c$, F.~Rotondo$^b$, A.~Rugliancich$^o$,
T.~Saito$^a$, Y.~Sunada$^k$, M.~Suzuki$^p$,
M.~Takahashi$^a$, L.A.~Tejedor$^n$,
P.~Vallania$^b$, C.~Vigorito$^b$,
T.~Yamamoto$^q$, T.~Yoshida$^p$,
for the CTA Consortium\footnote{for consortium list see PoS(ICRC2019)1177}\\
\scriptsize
\llap{$^a$} Institute for Cosmic Ray Research, University of Tokyo
5-1-5, Kashiwa-no-ha, Kashiwa, Chiba 277-8582, Japan\quad
\llap{$^b$} INFN Sezione di Torino and Universit\`{a} degli Studi di Torino
Via P. Giuria 1, 10125 Torino, Italy\quad
\llap{$^c$} INFN Sezione di Padova and Universit\`{a} degli Studi di Padova
Via Marzolo 8, 35131 Padova, Italy\quad
\llap{$^d$} Inst. de Astrof\'{i}sica de Canarias, and Universidad de La Laguna, Dpto. Astrof\'{i}sica
E-38200 La Laguna, Tenerife, Spain\quad
\llap{$^e$} Institut de F\'{i}sica d’Altes Energies (IFAE), The Barcelona Institute of Science and Technology
Campus UAB, 08193 Bellaterra (Barcelona), Spain\quad
\llap{$^f$} Aix Marseille Univ, CNRS/IN2P3, CPPM, Marseille, France
163 Avenue de Luminy, 13288 Marseille cedex 09, France\quad
\llap{$^g$} CIEMAT, Avda. Complutense 40
28040 Madrid, Spain\quad
\llap{$^h$} INFN Sezione di Bari and Universit\`{a} degli Studi di Bari
Via Orabona 4, 70124 Bari, Italy\quad
\llap{$^i$} Division of Physics and Astronomy, Graduate School of Science, Kyoto University
Sakyo-ku, Kyoto, 606-8502, Japan\quad
\llap{$^j$} Max-Planck-Institut f\"{u}r Physik
F\"{o}hringer Ring 6, 80805 M\"{u}nchen, Germany\quad
\llap{$^k$} Graduate School of Science and Engineering, Saitama University
255 Simo-Ohkubo, Sakura-ku, Saitama city, Saitama 338-8570, Japan\quad
\llap{$^l$} Institute for Space-Earth Environmental Research, Nagoya University
Chikusa-ku, Nagoya 464-8601, Japan\quad
\llap{$^m$} University of Siena and INFN
via Roma 56, 53100 Siena, Italy\quad
\llap{$^n$} Grupo de Altas Energ\'{i}as and UPARCOS, Universidad Complutense de Madrid
Av Complutense s/n, 28040 Madrid, Spain\quad
\llap{$^o$} INFN Sezione di Pisa
Largo Pontecorvo 3, 56217 Pisa, Italy\quad
\llap{$^p$} Faculty of Science, Ibaraki University
Mito, Ibaraki, 310-8512, Japan\quad
\llap{$^q$} Department of Physics, Konan University
Kobe, Hyogo, 658-8501, Japan
\scriptsize

        E-mail: \email{ssakurai@icrr.u-tokyo.ac.jp}}

\abstract{\scriptsize
The Cherenkov Telescope Array (CTA) represents the next generation of very high-energy gamma-ray observatory, which will provide broad coverage of gamma rays from 20 GeV to 300 TeV with unprecedented sensitivity.
CTA will employ three different sizes of telescopes, and the Large-Sized Telescopes (LSTs) of 23-m diameter dish will provide the sensitivity in the lowest energies down to 20 GeV.
The first LST prototype has been inaugurated in October 2018 at La Palma (Canary Islands, Spain) and has entered the commissioning phase.
The camera of the LST consists of 265 PMT modules.
Each module is equipped with seven high-quantum-efficiency Photomultiplier Tubes (PMTs), a slow control board, and a readout board.
Ensuring high uniformity and precise characterization of the camera is the key aspects leading to the best performance and low systematic uncertainty of the LST cameras.
Therefore, prior to the installation on site, we performed a quality check of all PMT modules.
Moreover, the absolute calibration of light throughput is essential to reconstruct the amount of light received by the telescope.
The amount of light is affected by the atmosphere, by the telescope optical system and camera, and can be calibrated using the ring-shaped images produced by cosmic-ray muons.
In this contribution, we will show the results of off-site quality control of PMT modules and on-site calibration using muon rings.
We will also highlight the status of the development of Silicon Photomultiplier modules that could be considered as a replacement of PMT modules for further improvement of the camera.
\scriptsize}

\FullConference{36th International Cosmic Ray Conference -ICRC2019-\\
		July 24th - August 1st, 2019\\
		Madison, WI, U.S.A.}

\begin{document}

\section{Introduction}
The Cherenkov Telescope Array (CTA) project is an ongoing imaging atmospheric Cherenkov telescope (IACT) project.
IACTs observe Cherenkov light emitted by gamma-ray induced extensive air showers and reconstruct the energy and direction of gamma rays arriving at the earth.
CTA will perform observations in the energy range between 20~GeV and 300~TeV with 10 times better flux sensitivity of current facilities.
For this purpose, CTA will employ three types of telescopes: Large-Sized Telescope (LST), Medium-Sized Telescope (MST), and Small-Sized Telescope (SST).
While the CTA northern site (La Palma, Spain) will host only LSTs and MSTs, the CTA southern site (Paranal, Chile) will host also SSTs.

Since the Cherenkov light flash produced by gamma rays of several tens of GeV is extremely faint, a large mirror is required to collect enough photons.
LST, the largest telescope of CTA, will be dedicated to the energy range from tens of GeV to several TeV \cite{LST}.

In order to capture fainter Cherenkov light, LST is equipped with a 23 m diameter split parabolic mirror and a high-sensitivity camera made by 1855 pixels.
Despite a mass of 120~tons, LST has been designed to point any direction in sky within 20~seconds, to allow the observation of transient phenomena such as Gamma-Ray Bursts (GRBs).

\begin{figure}

	\begin{minipage}[t]{0.5\hsize}
		\centering
		\includegraphics[width=\hsize]{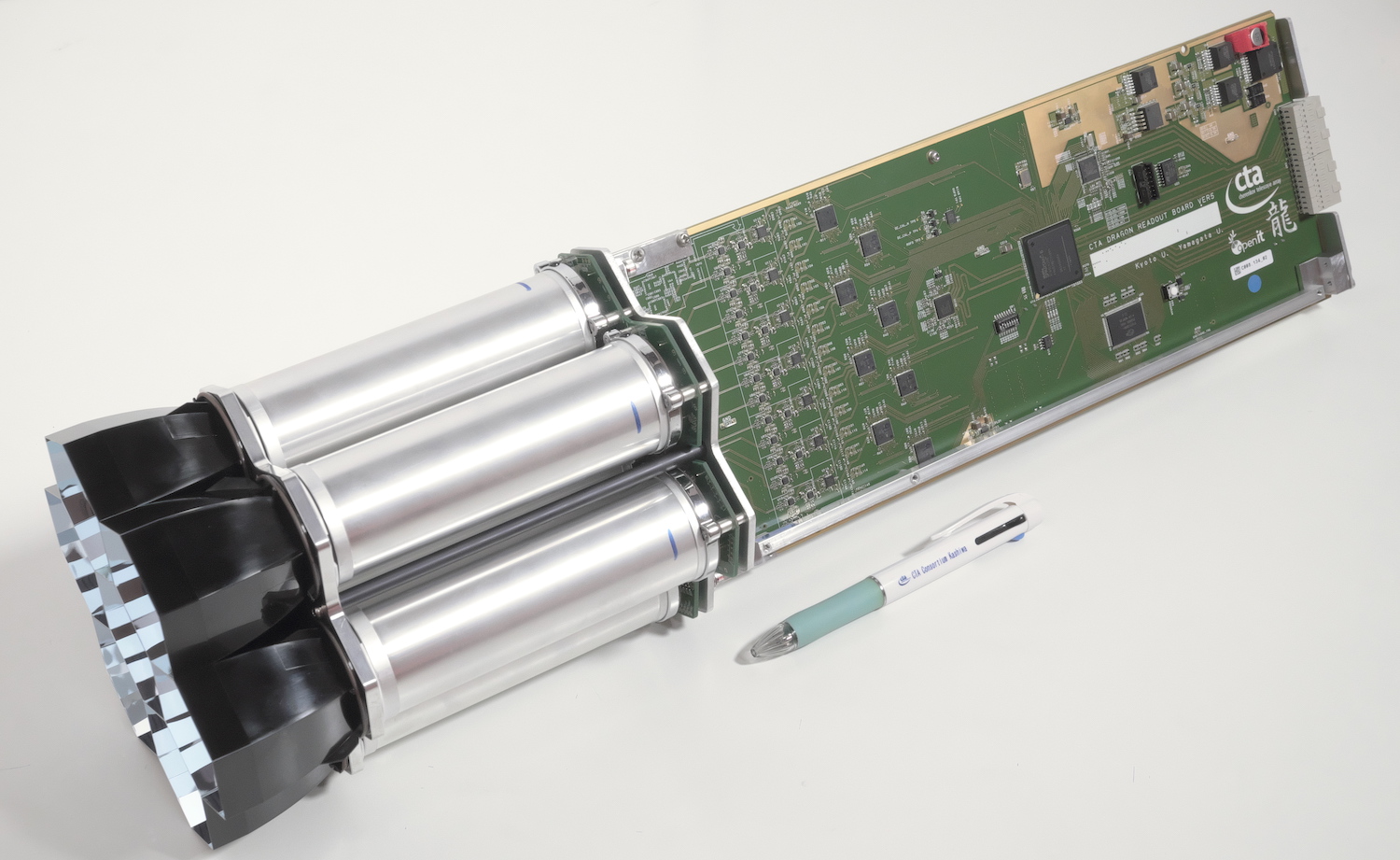}
		\subcaption{PMT module for the LST camera.}
	\label{pmodule}
	\end{minipage}
	\begin{minipage}[t]{0.5\hsize}
		\centering
		\includegraphics[width=\hsize]{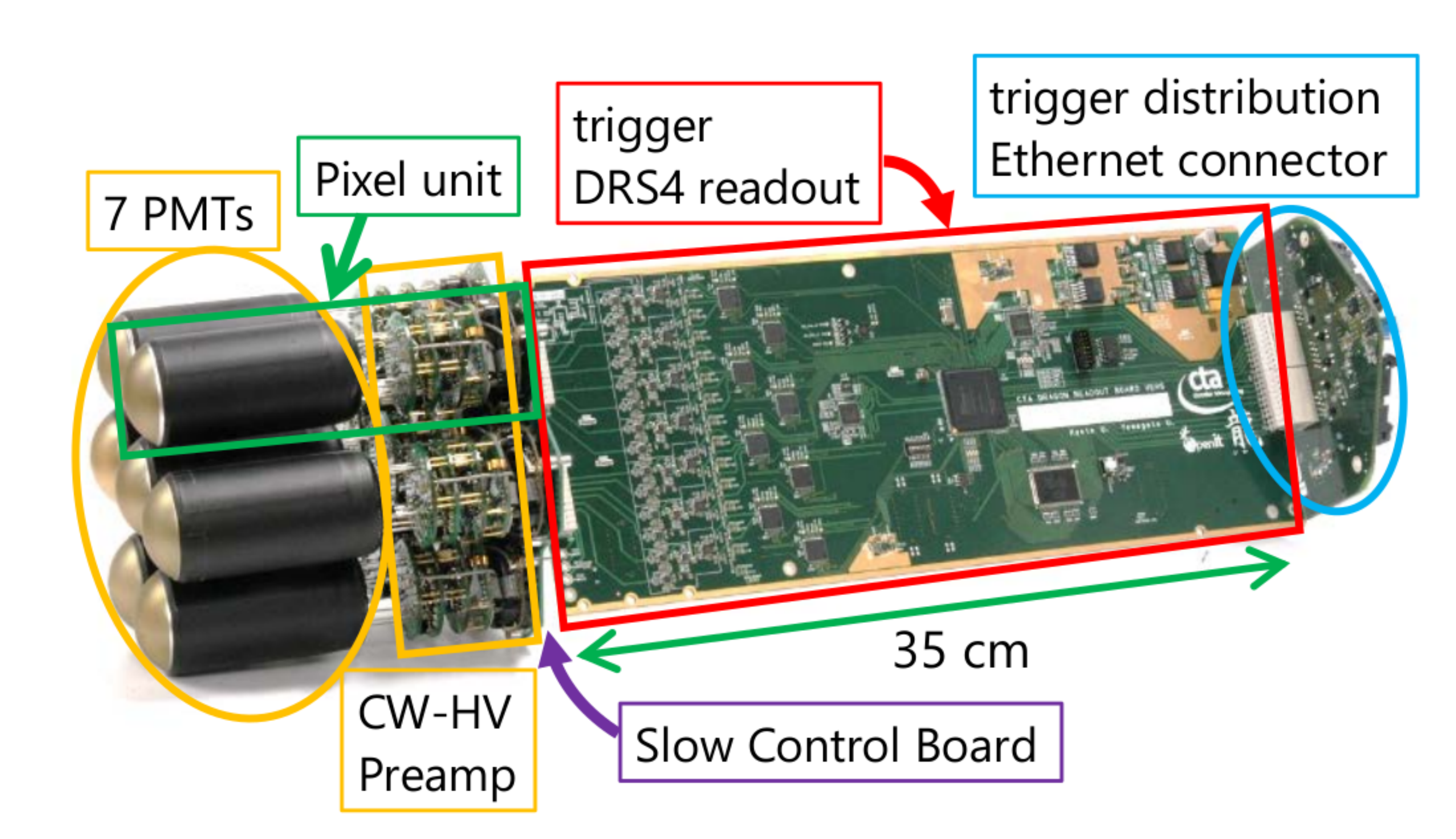}
		\subcaption{CTA-LST electronic chain \cite{DevPMTLST}}
		\label{fig:LST_PMT_01}
	\end{minipage}
	\caption{Overview of PMT module for the LST}

\end{figure}

The LST camera is composed by 265 "PMT modules" (shown in figure \ref{pmodule}). Each module hosts a signal readout board, a slow control board and 7 pixel units, each one composed by a Cockcroft Walton HV supplier, a preamplifier and a high quantum efficiency Photo-Multiplier Tube (PMT) with a light collector in front of it \cite{DevPMTLST}.

The LST1 was inaugurated in October 2018.
Currently, the first LST prototype is under commissioning operations.

In this contribution, we report the results of the quality inspection test of PMT module performed prior to the camera installation on the telescope, the current status of the commissioning operations, and the status of development of the SiPM-based module.

\section{PMT module and quality check test}

\subsection{PMT module}
The PMTs were developed by CTA-Japan and Max-Planck-Institut f\"{u}r Physik with Hamamatsu Photonics K.K.
The average peak quantum efficiency of the PMTs used for the first LST prototype reaches 40\% around 350~nm wavelength.
The light collector reduces the dead area between pixels in the camera and suppresses the off-axis light contamination.

Since the signal readout board reads out the differential outputs of seven PMTs with two different amplifications, eight ASICs for analog sampling called Domino Ring Sampler 4 (DRS4) are mounted.
This chip was developed by the Paul Scherrer Institut for the MEG experiment \cite{DRS} and enables high-speed analog waveform sampling up to 5~GHz and readout at 33~MHz.

Prior to installation, we  carried out a quality check measurement of the PMT modules at the Institute for Cosmic Ray Research in Japan and Instituto de Astrof\'{i}sica de Canarias on Tenerife, Canary Islands, Spain.

\subsection{Measurement system}
PMT module quality check was performed from 2016 to 2017.
We performed the measurement using a pulsed laser for checking the response of the PMTs and the signal readout board.
We measured the gain of the PMTs as a function of the high voltage (HV) and defined "nominal HV" the one corresponding to a gain of $4\times10^4$.
Single photoelectron measurement was performed to evaluate the signal-to-noise ratio with nominal HV, and then the linearity and the charge resolution of the signal up to 1000 photoelectrons were evaluated by changing the amount of light of the laser using neutral-density filters.

\begin{figure}
	\centering
	\includegraphics[width=0.8\hsize]{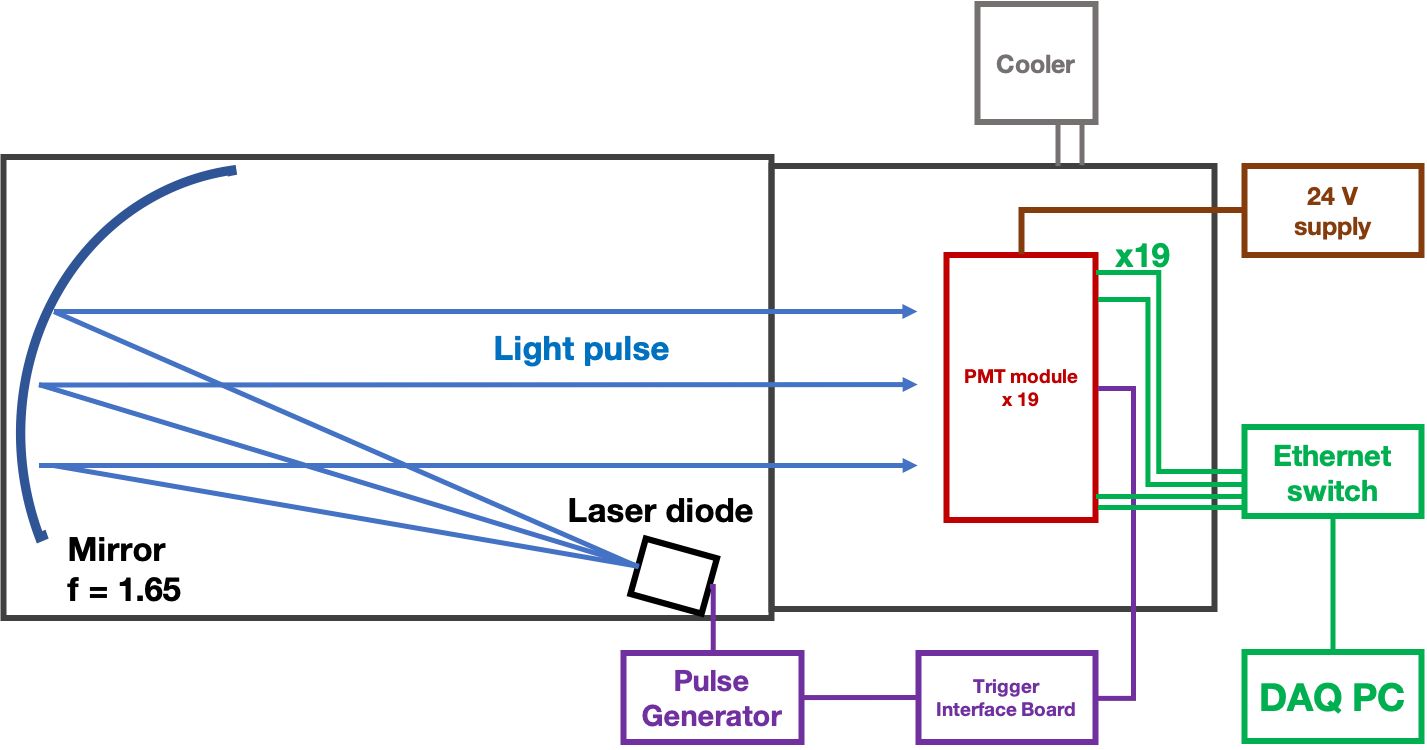}
	\caption{Mini camera  setup schematic view}
	\label{fsetup}
\end{figure}

We performed the tests with a setup hosting 19 PMT modules ("mini camera") inside a dark box (1.5 x 1.5 x 3.5 m), as shown in figure \ref{fsetup}.
The sensors were illuminated using a laser and a reflector.
The triggers, synchronized with the laser, were handled by the Trigger Interface Board (TIB), which is also used in the LST camera.

\subsection{Measurement result}

\begin{figure}

	\begin{minipage}[t]{0.5\hsize}
		\centering
		\includegraphics[width=\hsize]{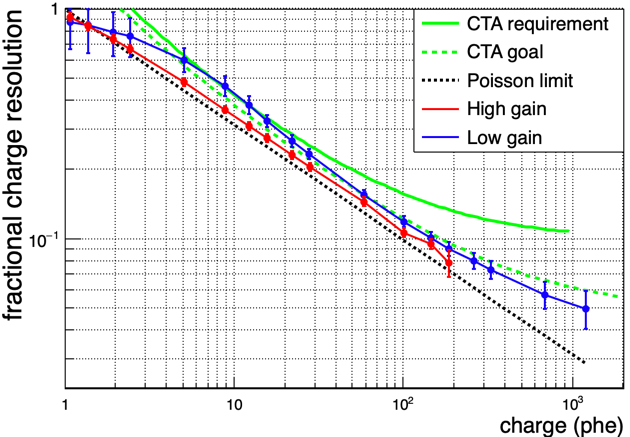}
		\subcaption{Charge resolutions of High and Low gain signals.}
	\label{fcharge}
	\end{minipage}
	\begin{minipage}[t]{0.5\hsize}
		\centering
		\includegraphics[width=\hsize]{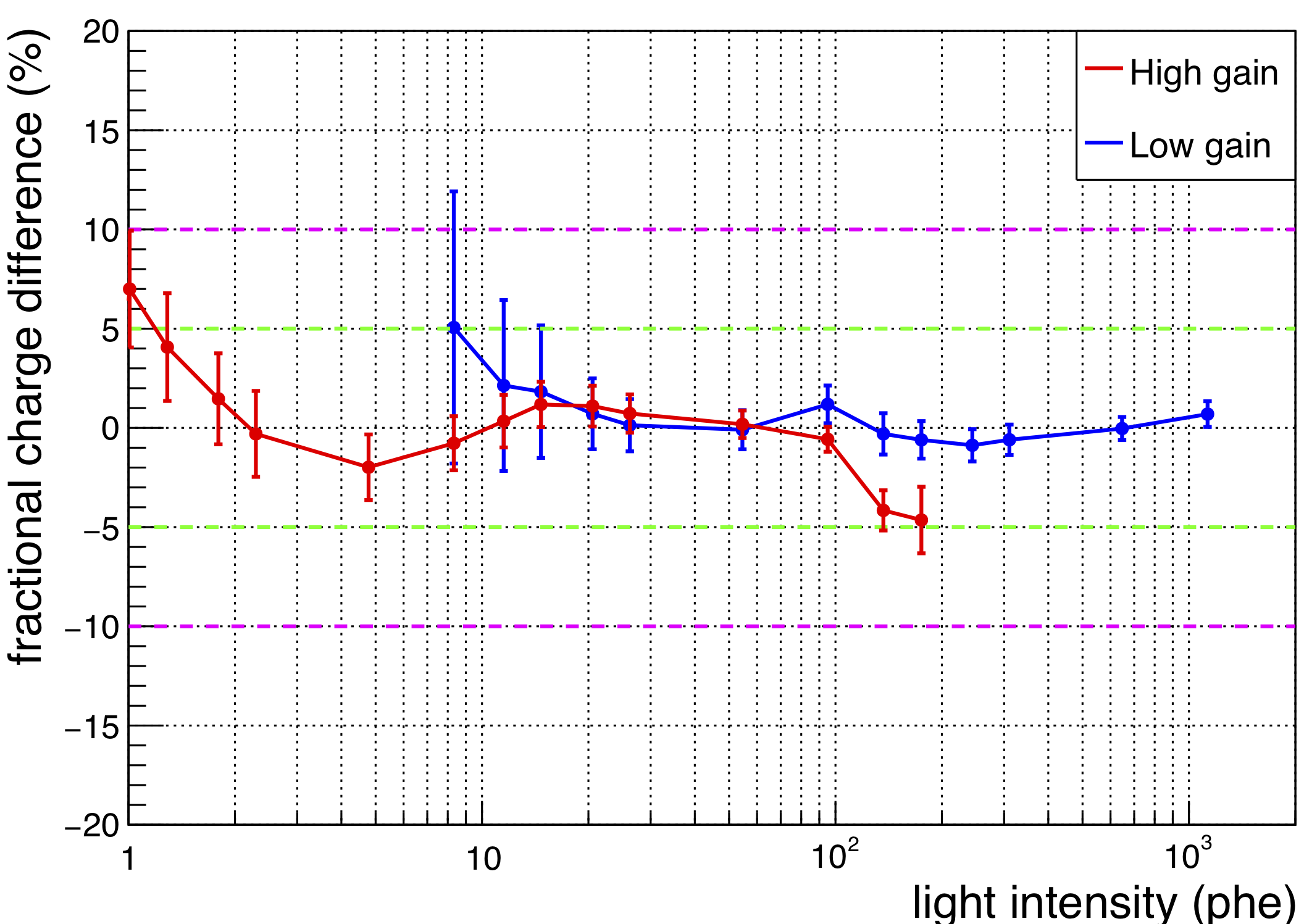}
		\subcaption{Linearity of the two channels. Additional lines are shown at 5\% and 10\% as a reference.}
		\label{flinearity}
	\end{minipage}
	\caption{Results of measurement using three orders of magnitude of light intensity }

\end{figure}

The result of the charge resolution measurement is shown in figure \ref{fcharge}.
The red points show fractional charge resolution of the high gain channel, blue points show fractional charge resolution of the low gain channel.
The green solid and dotted lines represent the CTA requirement and goal, respectively.
Both high and low gain reach the goal performance. 

 The result of linearity measurement is shown in figure \ref{flinearity}.
 Red points show a fraction of the charge difference from the laser intensity in high gain.
 Blue points show that in low gain.
 Measured points are inside 5\% difference in the photo-electron range from 1~p.e. to $10^3$~p.e.

\section{Muon analysis for the LST}

IACTs use camera calibration systems to estimate the conversion between the measured signal and the total number of received photons. This method, however, does not take into account effects of the optical system and the efficiency of the mirror of the telescope is not considered. To carry out an absolute calibration of light throughput, it is essential to analyze a signal of known nature, as for example the peculiar ring-shaped images produced by cosmic ray muons \cite{muons_cta}.
The LST of CTA will work with a stereo trigger that will allow to lower its energy threshold, but this will be also its capability of detecting muon events, usually triggering only one telescope.
In this work we pursued two objectives: prepare the analysis setup to evaluate the optical throughput of the system using single-telescope triggered muons and evaluate the stereo muon rate of the array of 4 LSTs.

\subsection{Simulation setup}

For the simulation of muon rings we used the simtelarray software \cite{simtelarray}. To save simulation time, we simulated muons triggered by a single telescope with a special configuration that allowed to trigger most of the simulated events. To test what is the capability of detecting well reconstructed muons with a stereo trigger, we simulated 10$^7$ muon events at 7 km height, with a Viewcone of 3.5 deg and up to an impact parameter of 150 m. For the analysis of muons rings we used the ctapipe software \cite{ctapipe}. An example of a muon analyzed with ctapipe is shown in Figure \ref{fig:muon}.

\subsection{Results}

For stereo rate, we identified 639 out of the 10$^7$ muons simulated. Assuming that we have a flux of F = 200 $\mu$ m$^{-2}$s$^{-1}$sr$^{-1}$ at 7 km altitude \cite{pdg_muons} and using the solid angle and collection area corresponding to our simulations, we find a rate of 10 Hz for stereo well-reconstructed muons for four LSTs. This rate may be enough to monitor the optical efficiency of the system without having to take dedicated muon runs with a single-telescope trigger, with the subsequent loss of observation time.

\begin{figure}

	\begin{minipage}[t]{0.5\hsize}
		\centering
		\includegraphics[width=\hsize]{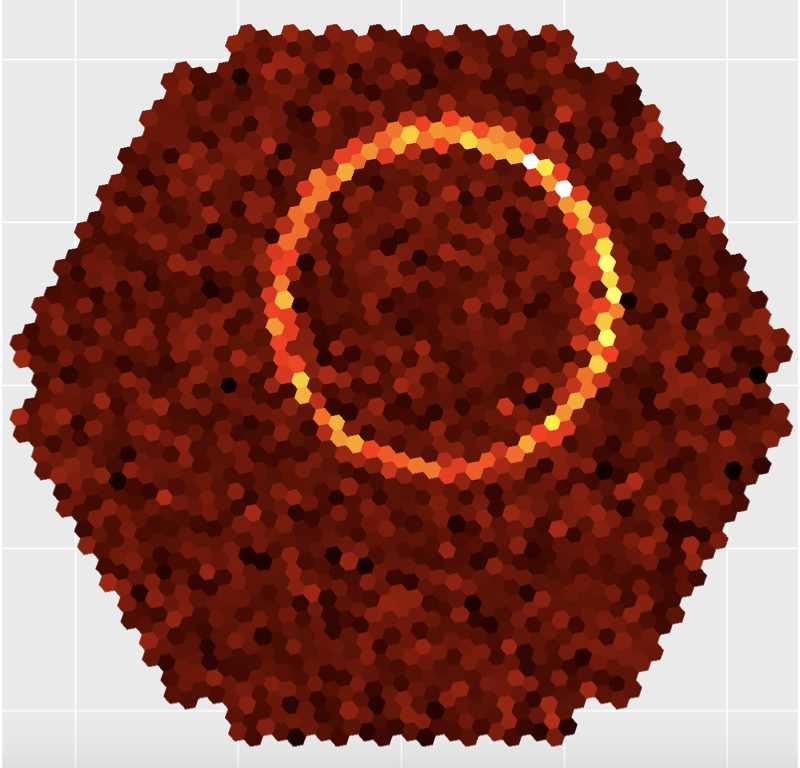}
		\subcaption{Example of a Muon simulated and analyzed using the CTA simulation software.}
	\label{fig:muon}
	\end{minipage}
	\begin{minipage}[t]{0.5\hsize}
		\centering
		\includegraphics[width=\hsize]{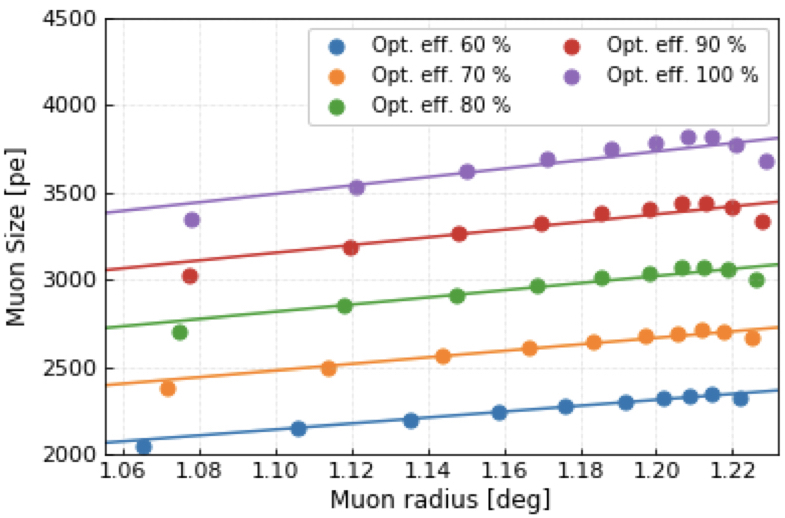}
		\subcaption{Muon ring width as a function of ring size.}
		\label{fig:width_vs_size}
	\end{minipage}
	\caption{Muon analysis images}

\end{figure}

In Figure \ref{fig:width_vs_size}, we show a plot of the muon ring width as a function of the size (number of photoelectrons) of the image. We will compare the results of this simulation with those of the real data in the framework of the commissioning of the first LST, located in the island of la Palma.

\section{Development of a SiPM-based Camera}

While the PMT technology is mature and reliable, Silicon Photomultipliers (SiPMs) are more and more emerging as promising competitors in many low-light level detection applications. Concerning IACT experiments, their higher quantum efficiency can lower the telescope energy threshold, thus increasing the scientific potential of the instrument; for example, lowering it allows the detector to be sensitive to fainter sources and, since the $\gamma$-ray absorption increases as $ E_{\gamma} $ increases, to access deeper regions in the Universe. Moreover they are capable to tolerate high illumination level (e.g. moonlight): this can significantly increase the Duty Cycle of these facilities, which, for example, can be crucial for the study of transient phenomena and for time-demanding campaigns. In addition they work at lower operating voltage, which simplifies the design an the operation of the supply electronics, and they are more robust. On the other hand they have few disadvantages: while their higher dark count rate is not a real cause of concern, since it is anyway lower than the night sky background, more attention must be paid to the cross talk phenomenon, which is a peculiarity of these sensors (nonetheless many efforts are done to significantly lowering it). It is also necessary to point out their higher device capacitance, which can worsen the signal to noise ratio when compared to traditional PMTs, and their less narrow PDE spectrum, that makes them more sensitive to the Night Sky Background.
\par
The first Geiger Mode Avalanche Photodiode Cherenkov Telescope (FACT) \cite{FACT}, in operation since 2011, has clearly shown the reliability of silicon detector for this application. Moreover the advantage of using these sensors for small size telescopes has been proven, since in the CTA framework SiPMs are already chosen as photodetectors for the SSTs and for the dual mirror medium-size telescopes \cite{Biteau}. We are pursuing this R\&D in order to find out if the SiPM solution is advantageous (or at least comparable to PMTs) also for the LSTs.

\subsection{Project}

\begin{figure}

	\begin{minipage}[t]{0.5\hsize}
		\centering
		\includegraphics[width=\hsize]{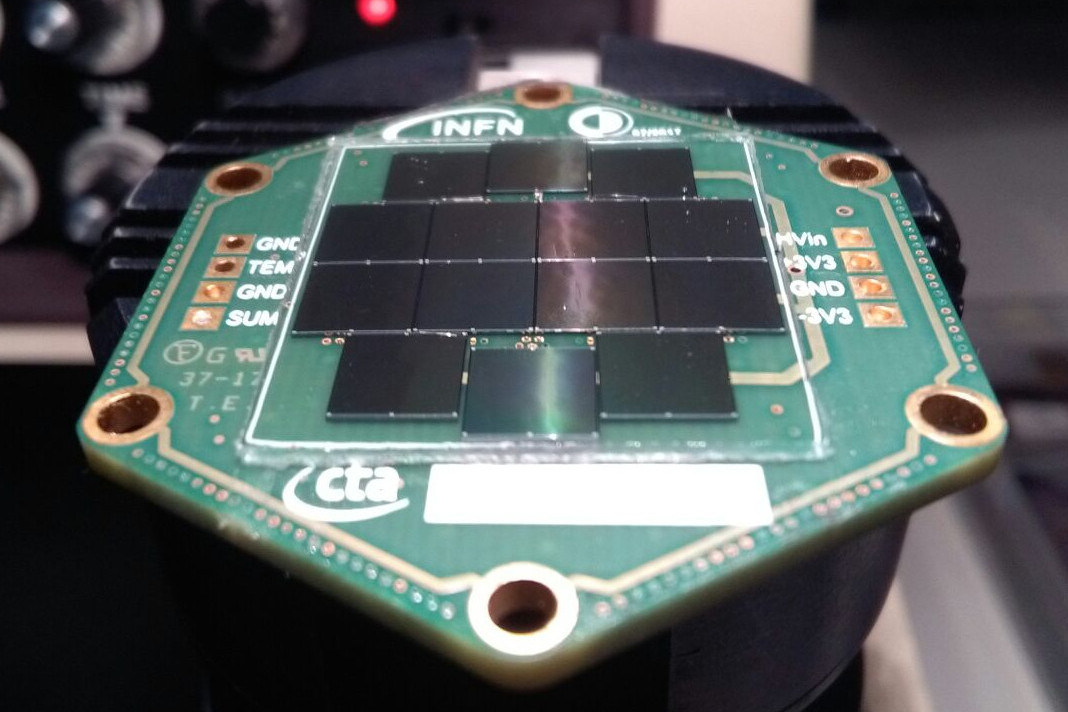}
		\subcaption{SiPM Pixel Board Prototype}
	\label{fig:PB_Prototype_03_cut}
	\end{minipage}
	\begin{minipage}[t]{0.5\hsize}
		\centering
		\includegraphics[width=\hsize]{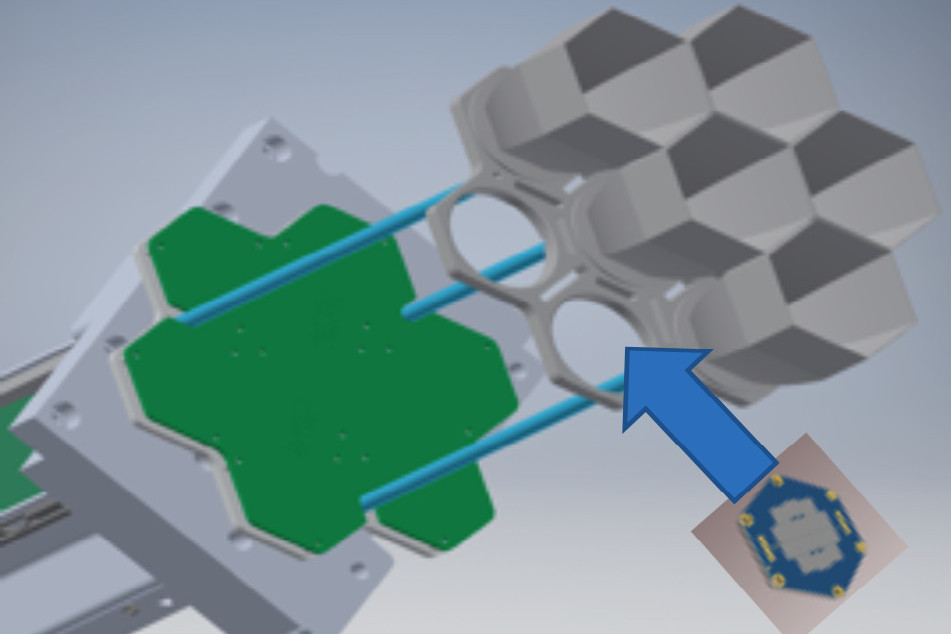}
		\subcaption{Rendering of the SiPM-based camera prototype concept}
		\label{fig:PB_Structure_01}
	\end{minipage}


	\begin{minipage}[t]{0.5\hsize}
		\centering
		\includegraphics[width=\hsize]{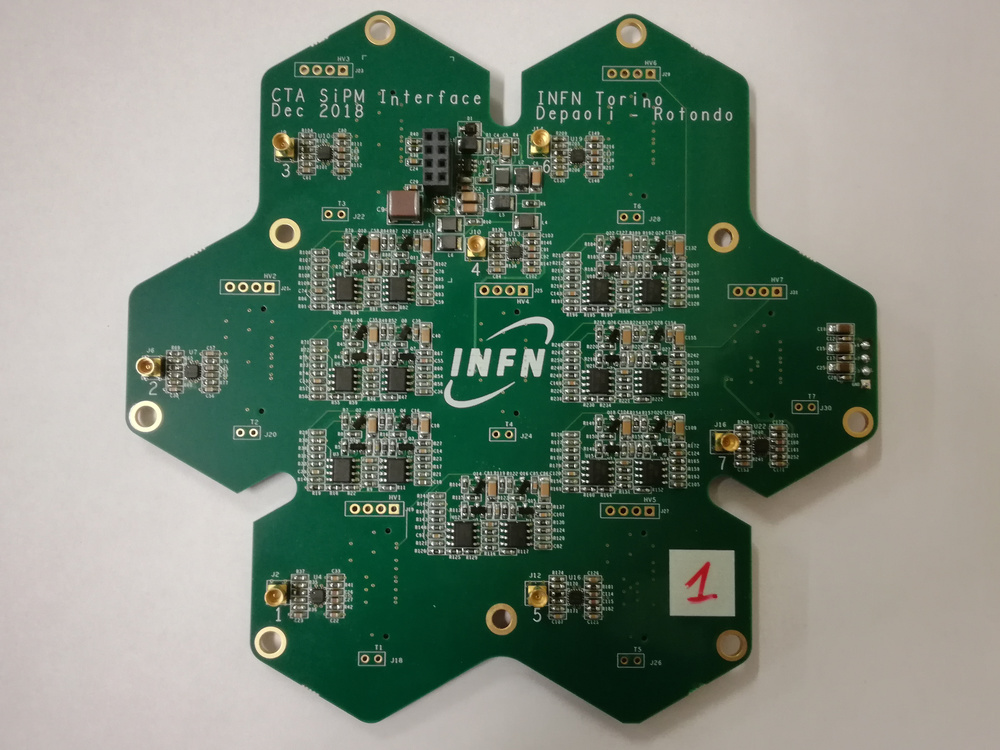}
		\subcaption{Interface Board Prototype}
	\label{fig:IB_Prototype_01}
	\end{minipage}
	\begin{minipage}[t]{0.5\hsize}
		\centering
		\includegraphics[width=\hsize]{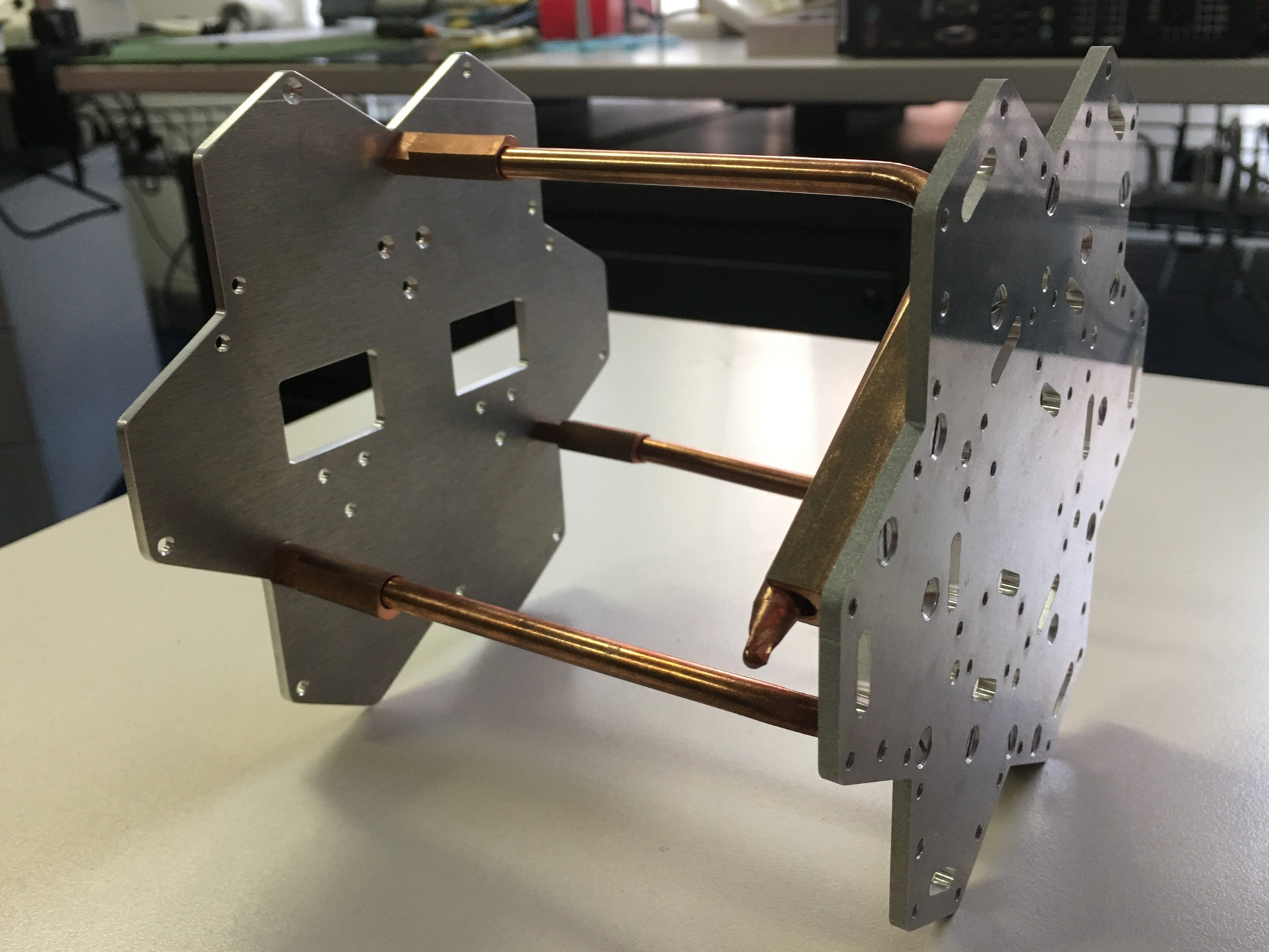}
		\subcaption{Mechanical structure with the heat pipe}
		\label{fig:Mec_Structure_01}
	\end{minipage}
	\caption{Interfaces for SiPM-based readout system}

\end{figure}

In order to contain the cost, most of the existing camera readout is kept \cite{SiPMRD}; referring to Figure \ref{fig:LST_PMT_01}, the idea is to use all the electronic chain untill the Slow Control Board (SCB), replacing the PMT Pixel Units 
substituting each 1.5-inch PMT with a matrix of 14 SiPMs; the sensors chosen for this R\&D project are the Fondazione Bruno Kessler (FBK) NUV HD3-2 with an area of 6 mm $\times$ 6 mm, developed in order to have the peak of the Photon Detection Efficiency close to the peak of the Cherenkov Spectrum. This matrix is hosted on an electronic module (called Pixel Board) which sums and shapes the SiPM signals.
The concept idea is shown in Fig. \ref{fig:PB_Structure_01}. The use of SiPMs on large area detectors is challenging and similar efforts are ongoing within the MAGIC collaboration \cite{SiPMRD}.
\par
The connection between the Pixel Board and the Slow Control Board is made by an ``Interface Board'', which substitutes the Cockcroft-Walton Preamplifier; the first prototype is shown in Figure \ref{fig:IB_Prototype_01}. This board hosts a low-noise DC-DC converter to bias the SiPMs; the supply voltage is set independently for each channel by a Digital to Analog Converted housed in the SCB. Moreover, in order to have a closed negative feedback, it measures the voltage on the Pixels and the current that flows in the channels: this values are then read by an Analog to Digital Converter placed in the Slow Control Boards. Each channel of the Interface Board has an Single-ended to Differential Converter, to adapt the output signal of the Pixel Boards for the remaining part of the electronic chain.
Since the SiPM response depends on their temperature, the Pixel Boards are connected to a cooled aluminum plate, which maintains the temperature at about 15$^\circ$; in addition a digital thermometer measures the temperature on the sensors, used to correct the bias voltage.
In Figure \ref{fig:Mec_Structure_01} is shown the structure designed to host the SiPM boards.
\par
The performance of the SiPM module will be compared to the one of the standard PMT.

\section{Summary}
The first LST prototype started commissioning operations in October 2018.
The PMT modules for the LST camera passed several quality checks.
And some parameters tuning (e.g. timing delays of the PMT signal for the event trigger and nominal HV) are being performed.
The LST camera will be ready for the science operation soon.

The absolute calibration using muon rings will be performed soon using the real observation data and it looks promising.
We will be able to get the factor to convert from the signal intensity to the number of photons.

There are ongoing R\&D activities to explore possible solutions to use SiPMs in the LST camera.

\clearpage
\begin{thebibliography}{99}

\bibitem{LST} D. Mazin, J. Cortina, M. Teshima, and the CTA Consortium Large size telescope report AIP Conference Proceedings 1792, 080001 2017

\bibitem{DevPMTLST} Shu Masuda et al., ``Development of the photomultiplier tube readout system for the first Large-Sized Telescope of the Cherenkov Telescope Array'', arXiv:1509.00548v1, 2 Sep 2015


\bibitem{DRS} S. Ritt, R. Dinapoli, and U. Hartmann. Application of the DRS chip for fast waveform digitizing. Nuclear Instruments and Methods in Physics Research A, 623, 486–488, November 2010

\bibitem{muons_cta} The CTA Consortium, USING MUON RINGS FOR THE OPTICAL THROUGHPUT CALIBRATION OF THE CHERENKOV TELESCOPE ARRAY - PART I, in preparation

\bibitem{simtelarray} Bernl\"or, K., Simulation of Imaging Atmospheric Cherenkov Telescopes with CORSIKA and sim\_telarray, APh, 30, 3 (2008).


\bibitem{ctapipe} K. Kosack for the CTA Consortium, ctapipe: A Low-level Data Processing Framework for CTA, this proceedings, id 717 (2019).


\bibitem{pdg_muons} Patrignani, C. et al., Review of Particle Physics - PDG, Chin.Phys. C40 (2016) no.10, 100001

\bibitem{FACT} A. Biland et al., “Calibration and performance of the photon sensor response of FACT - The First G-APD Cherenkov telescope”, arXiv:1403.5747v2, 30 Jul 2014

\bibitem{Biteau}  Jonathan Biteau et al., “Performance of Silicon Photomultipliers for the Dual-Mirror MediumSized Telescopes of the Cherenkov Telescope Array”, arXiv:1508.06245v1, 25 Aug 2015

\bibitem{SiPMRD} Riccardo Rando et al., ``Silicon Photomultiplier Research and Development Studies for the Large Size Telescope of the Cherenkov Telescope Array'', arXiv:1508.07120v1, 28 Aug 2015


\end{thebibliography}
\end{document}